\documentclass[a4paper,12pt]{article}
\usepackage{epsfig}
\usepackage{color}
\usepackage[numbers,sort&compress]{natbib}

\newlength{\dinwidth}
\newlength{\dinmargin}
\newcommand{\spur}[1]{\not\! #1 \,}

\begin{document}

\title{Reevaluating  R-parity Violating
Supersymmetry Effects in  $B_s^0-\bar{B}_s^0$ Mixing}
\author{ Ru-Min Wang\thanks{E-mail: ruminwang@gmail.com},~
  Yuan-Guo Xu\thanks{E-mail: yuangx@iopp.ccnu.edu.cn},~ Mo-Lin Liu\thanks{E-mail: mlliu@mail2.xytc.edu.cn},~ Bing-Zhong Li\thanks{E-mail: libingzhong08@yahoo.cn}
  \\
{\scriptsize \it College of Physics and Electronic Engineering,
Xinyang Normal University,
 Xinyang, Henan 464000, China
}
 }
 \maketitle

\begin{abstract}
Recently, the CDF and D{\O} collaborations have claimed that the CP
violating phase in  $B_s^0-\bar{B}_s^0$ mixing is large, which is
contrary to the expectations in the Standard Model. Such a large
phase suggests New Physics contributions to $B_s^0-\bar{B}_s^0$
mixing. Motivated by this, we reevaluate the constraints on R-parity
violating contributions, including baryon number violating couplings
not considered before, to the mixing mass matrix element $M^s_{12}$
from the recent measurements of
 $B_s^0-\bar{B}_s^0$ mixing. We show that present data allow us
to put quite strong constraints on both the magnitudes and the weak
phases of the R-parity violating parameters. Some of these  bounds
are better than the existing ones, and some bounds are obtained for
the first time. Near future experiments at the Tevatron, the LHC and
B-factories can shrink or reveal the relevant parameter spaces of
the R-parity violating couplings.

\end{abstract}

 \noindent {\bf Keywords:}  B Physics, Neutral
meson mixing, Supersymmetry,  R-parity violating\\

\noindent {\bf PACS Numbers:  12.60.Jv, 11.30.Er, 12.15.Mm,
14.40.Nd}

\newpage

\section{Introduction}
Mixing phenomena in heavy bosons system is considered as an
important test of the Standard Model (SM) and a probe for New
Physics (NP) beyond the SM. Recently, the large CP violating phase
$\phi_s^{J/\psi\phi}$ associated with $B_s^0-\bar{B}_s^0$ mixing
has been obtained from the time-dependent angular analysis of
flavor-tagged $B^0_s\to J/\psi\phi$ decay by the CDF and D{\O}
collaborations \cite{Aaltonen:2007he,:2008fj,Tonelli:2008ey}. More
recently,  the D{\O} collaboration has announced evidence for a
charge asymmetry in the number of like-sign dimuon events
\cite{Abazov:2010hv}, which can be interpreted as further evidence
for large CP violation in $B_s^0-\bar{B}_s^0$ mixing. The CP
violating phase measured by both CDF and D{\O} is \cite{exphis}
\begin{eqnarray}
\phi^{J/\psi\phi}_s &\in& [0.54, 1.18]\cup [1.94,2.60]~ (\mbox{at}~ 68\%~ \mbox{C.L.}), \nonumber\\
\phi^{J/\psi\phi}_s &\in& [0.20, 2.84]~ (\mbox{at}~
95\%~\mbox{C.L.}).
\end{eqnarray}
However,  the CP violating phase is predicted  precisely to be
 small in the SM
\cite{Bona:2009tn,Lenz:2006hd,Silvestrini:2008zza,Bona:2008jn,Barberio:2008fa},
$\phi^{J/\psi\phi,\mbox{\scriptsize
SM}}_s=2\beta_s^{\mbox{\scriptsize
 SM}}\equiv
2\mbox{arg}\left(-\frac{V_{ts}V^*_{tb}}{V_{cs}V^*_{cb}}\right)\approx
0.04$. Current experimental data  of the  CP violating phase deviate
about $2\sigma$ from its SM value, which indicates that there are
possible large NP contributions to the phase, i.e.,
$\phi^{J/\psi\phi}_s=\phi^{J/\psi\phi,\mbox{\scriptsize
SM}}_s+\phi^{\mbox{\scriptsize  NP}}_s$.

In general, the relevant CP violating phase between the
$B^0_s-\bar{B}^0_s$ amplitude and the amplitudes of the subsequent
$B^0_s$ and $\bar{B}^0_s$ decay to a common final state  could be
expressed as \cite{Lenz:2007nk}
\begin{eqnarray}
\phi_s=\mbox{arg}\left(-\frac{M^s_{12}}{\Gamma^s_{12}}\right),
%0803.0659 =2(\beta_s-\phi_{B_s})$
\end{eqnarray}
where $M^s_{12}$ is the off-diagonal element of the $\Delta B=2$
mass matrix, and $\Gamma^s_{12}$ is the off-diagonal element of the
decay matrix. The SM prediction for this phase is tiny,
$\phi^{\mbox{\scriptsize  SM}}_s\approx0.004$ \cite{Lenz:2006hd}.
The same additional contribution $\phi^{\mbox{\scriptsize  NP}}_s$
due to NP would change this observed phase,  i.e.,
$\phi_s=\phi^{\mbox{\scriptsize SM}}_s+\phi^{\mbox{\scriptsize
NP}}_s$.

The current experimental precision does not allow these small
CP-violating phases $\phi^{J/\psi\phi,\mbox{\scriptsize  SM}}_s$ and
$\phi^{\mbox{\scriptsize  SM}}_s$ to be resolved. In case of sizable
NP contributions,
 the following
approximation is used:
$\phi^{J/\psi\phi}_s\approx\phi_s\approx\phi^{\mbox{\scriptsize
NP}}_s$.
% Such a large phase can be explained by NP contributions to both $M^s_{12}$ and $\Gamma^s_{12}$.
In order to explain the large CP asymmetry, the NP contributions to
$M^s_{12}$ and $\Gamma^s_{12}$ have already been widely studied in
recent works (for example, see Refs.
\cite{Buras:2010mh&Dobrescu:2010rh,Chen:2010aq&Chen:2010wv,Kubo:2010mh&Kawashima:2009jv,
Eberhardt:2010bm&Bobrowski:2009ng,Parry:2010ce&Dutta:2009hj,
Chang:2009tx&Deshpande:2010hy,Bauer:2010dg&Dighe:2010nj,Bai:2010kf&Choudhury:2010ya&King:2010np}).

As one of the most promising candidates for NP, Supersymmetry (SUSY)
\cite{Weinberg:1981wj,SUSY}, in both its R-parity conserving and its
R-parity Violating ($\spur{R}_p$) incarnations, is extensively
studied. In SUSY without R-parity, the following $\spur{R}_p$
superpotential are also allowed \cite{Weinberg:1981wj}
\begin{eqnarray}
\mathcal{W}_{\spur{R_p}}=\mu_i\hat{L}_i\hat{H}_u+\frac{1}{2}
\lambda_{[ij]k}\hat{L}_i\hat{L}_j\hat{E}^c_k+\lambda'_{ijk}\hat{L}_i\hat{Q}_j\hat{D}^c_k+\frac{1}{2}
\lambda''_{i[jk]}\hat{U}^c_i\hat{D}^c_j\hat{D}^c_k, \label{rpv}
\end{eqnarray}
where the first three terms violate lepton number, and the last term
violates baryon number.
$\spur{R}_p$ SUSY effects in neutral meson mixing have been
extensively discussed in the literatures (for example, Refs.
\cite{Nandi:2006qe,Saha:2003tq,Kundu:2004cv,Bhattacharyya:1998be,deCarlos:1996yh,Guetta:1998uk}).
In this paper, we focus on  the lepton number violating ($\spur{L}$)
and baryon number violating ($\spur{B}$) contributions to $M^s_{12}$
in SUSY model without R-parity.
Using the latest experimental data  of  $B_s^0-\bar{B}_s^0$ mixing,
we systematically evaluate the constraints on  relevant $\spur{L}$
and $\spur{B}$ couplings. We improve the bounds on relevant
$\spur{L}$ couplings from  current relevant data of
$B_s^0-\bar{B}_s^0$ mixing.  We explore the $\spur{B}$ coupling
effects in  $B_s^0-\bar{B}_s^0$ mixing for the first time. We find
our bounds on some $\spur{B}$ coupling products from
$B_s^0-\bar{B}_s^0$ mixing are better than ones from relevant
decays.  Moreover, the bounds on some $\spur{B}$ coupling products
are derived for the first time.

This paper is organized as follows. We briefly review the
theoretical basis for $B^0_s-\bar{B}^0_s$ mixing in Section 2, then
we deal with the numerical results in Section 3. In Section 4, we
summarize and conclude.

\section{Theoretical input for $B^0_s-\bar{B}^0_s$ mixing}

The most general $\Delta B=\Delta S=2$ process is described by the
effective Hamiltonian \cite{Becirevic:2001jj}
\begin{eqnarray}
\mathcal{H}_{eff}(\Delta B=\Delta
S=2)=\sum^5_{i=1}C_iQ_i+\sum^3_{i=1}\widetilde{C}_i\widetilde{Q}_i+h.c.,\label{Eq.Heff}
\end{eqnarray}
with
\begin{eqnarray}
Q_1&=&(\bar{s}\gamma^\mu P_Lb)_1(\bar{s}\gamma_\mu P_Lb)_1,\\
Q_2&=&(\bar{s} P_Lb)_1(\bar{s} P_Lb)_1,\\
Q_3&=&(\bar{s} P_Lb)_8(\bar{s} P_Lb)_8,\\
Q_4&=&(\bar{s} P_Lb)_1(\bar{s} P_Rb)_1,\\
Q_5&=&(\bar{s} P_Lb)_8(\bar{s} P_Rb)_8, \label{Q}
\end{eqnarray}
where $P_{L(R)}=(1-(+)\gamma_5)/2$ and the operators
$\widetilde{Q}_{1,2,3}$ are obtained from $Q_{1,2,3}$ by the
exchange $L\leftrightarrow R$. The hadronic matrix elements, taking
into account for renormalization effects, are defined as follows
\begin{eqnarray}
\langle \bar{B}^0_s|
Q_1(\mu)|B^0_s\rangle&=&\frac{2}{3}m^2_{B_s}f^2_{B_s}B^{(s)}_1(\mu),\\
\langle \bar{B}^0_s|
Q_2(\mu)|B^0_s\rangle&=&-\frac{5}{12}m^2_{B_s}f^2_{B_s}S_{B_s}B^{(s)}_2(\mu),\\
\langle \bar{B}^0_s|
Q_3(\mu)|B^0_s\rangle&=&\frac{1}{12}m^2_{B_s}f^2_{B_s}S_{B_s}B^{(s)}_3(\mu),\\
\langle \bar{B}^0_s|
Q_4(\mu)|B^0_s\rangle&=&\frac{1}{2}m^2_{B_s}f^2_{B_s}S_{B_s}B^{(s)}_4(\mu),\\
\langle \bar{B}^0_s|
Q_5(\mu)|B^0_s\rangle&=&\frac{1}{6}m^2_{B_s}f^2_{B_s}S_{B_s}B^{(s)}_5(\mu),\label{EM}
\end{eqnarray}
with
$S_{B_s}=\left(\frac{m_{B_s}}{\overline{m}_b(m_b)+\overline{m}_s(m_b)}\right)^2$.
The $B$-parameters  given in Table \ref{INPUT} have been taken from
Ref. \cite{Bparameter}.

The Wilson coefficients $C_i$ receive contributions from both the SM
and NP. In SUSY without R-parity, we only consider the $\spur{R}_p$
NP effects for the Wilson coefficients, $i.e.$, $C_i\equiv
C^{SM}_i+C^{\spur{R}_p}_i$.
\begin{figure}[t]
\begin{center}
\includegraphics[scale=1.1]{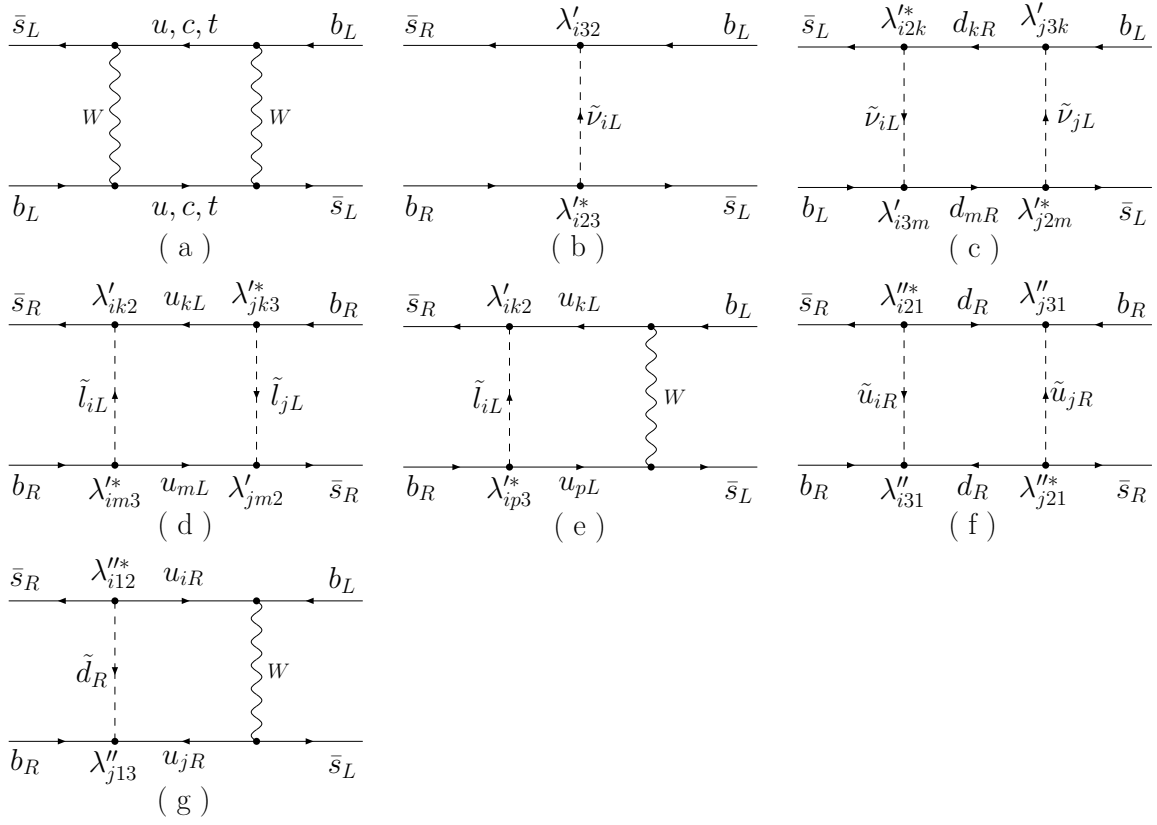}
\end{center}
\vspace{-0.9cm}
 \caption{ SM diagram (a), $\spur{L}$ diagrams (b-e) and $\spur{B}$ diagrams (f-g)
  which give the contributions  to $B^0_s-\bar{B}^0_s$ mixing.}
 \label{RPVdiagram}
\end{figure}
In the SM, the $t-W$ box diagram shown in Fig. \ref{RPVdiagram}(a)
generates only contribution to the operator $Q_1$, and the
corresponding Wilson coefficient $C^{SM}_1$ at the $m_b$ scale is
\cite{Buchalla:1995vs}
\begin{eqnarray}
C^{SM}_1(m_{b})=\frac{G_F^2}{4
\pi^2}m_W^2(V_{ts}V^*_{tb})^2\eta_{2B}S_0(x_t)[\alpha_s(m_b)]^{-6/23}
\left[1+\frac{\alpha_s(m_b)}{4\pi}J_5\right],
\end{eqnarray}
where $x_t=m^2_t/m^2_W$ and $\eta_{2B}$ is the QCD correction.

Now we turn to the $\spur{R}_p$ SUSY contributions to
$B^0_s-\bar{B}^0_s$ mixing. In the most general superpotential of
the minimal supersymmetric SM, there are new contributions to
$B^0_s-\bar{B}^0_s$ mixing from $\spur{R}_p$ couplings, and the
corresponding Wilson coefficients can be obtained from the
$\spur{R}_p$ superpotential given in Eq. (\ref{rpv})
\begin{eqnarray}
C^{\spur{R}_p}_1&=&C^{\lambda'}_1,\nonumber\\
C^{\spur{R}_p}_4&=&C^{TL}_4+C^{\lambda'W}_4+C^{\lambda''W}_4,\nonumber\\
C^{\spur{R}_p}_5&=&C^{\lambda'}_5+C^{\lambda''W}_5,\nonumber\\
\widetilde{C}^{\spur{R}_p}_1&=&\widetilde{C}^{\lambda'}_1+\widetilde{C}^{\lambda''}_1.\label{Hrpv}
\end{eqnarray}
Contributions to Wilson coefficients in Eq. (\ref{Hrpv})  come from
a variety of different classes of $\spur{R}_p$ diagrams, some of
which are shown in Fig. \ref{RPVdiagram}. One must add the diagrams
as shown in Fig. \ref{RPVdiagram}(c,d,f) by exchange of internal
fermions $\leftrightarrow$ corresponding internal sfermions.
$C^{TL}_4$ denotes the only tree level diagram with the exchange of
a sneutrino $\tilde{\nu}_i$ and two $\lambda'$ couplings, which is
shown in Fig. \ref{RPVdiagram}(b). The $\lambda'$ box diagrams such
as Fig. \ref{RPVdiagram}(c) give a contribution to
$C^{\lambda'}_1,\widetilde{C}^{\lambda'}_1$ and $C^{\lambda'}_5$.
The $\lambda'$ box diagrams such as Fig. \ref{RPVdiagram}(d) only
give a contribution to $\widetilde{C}^{\lambda'}_1$. The
$\lambda'-W$ box diagrams such as Fig. \ref{RPVdiagram}(e) with one
internal $W$ boson and one internal slepton give a contribution to
$C^{\lambda'W}_4$. The $\lambda''$ box diagrams such as Fig.
\ref{RPVdiagram}(f) give a contribution to
$\widetilde{C}^{\lambda''}_1$.  As
 shown in  Fig. \ref{RPVdiagram}(g), the $\lambda''-W$ box diagram gives a contribution to both $C^{\lambda''W}_4$ and
$C^{\lambda''W}_5$.
The coefficients in Eq. (\ref{Hrpv}) are given by
\cite{deCarlos:1996yh,Saha:2003tq} {\small
\begin{eqnarray}
C^{\lambda'}_1&=&\frac{1}{64\pi^2}\sum_{i,j,k,m}\lambda'^*_{i2k}\lambda'_{j3k}\lambda'^*_{j2m}\lambda'_{i3m}
\left[I_4(m^2_{\tilde{\nu}_i},m^2_{\tilde{\nu}_j},m^2_{d_k},m^2_{d_m})+I_4(m^2_{\nu_i},m^2_{\nu_j},m^2_{\tilde{d}^k_R},m^2_{\tilde{d}^m_R})\right],\\
\widetilde{C}^{\lambda'}_1&=&\frac{1}{64\pi^2}\sum_{i,j,k,m}\lambda'_{ik2}\lambda'^*_{jk3}\lambda'_{jm2}\lambda'^*_{im3}
\left[I_4(m^2_{\tilde{\nu}_i},m^2_{\tilde{\nu}_j},m^2_{d_k},m^2_{d_m})+I_4(m^2_{\nu_i},m^2_{\nu_j},m^2_{\tilde{d}^k_L},m^2_{\tilde{d}^m_L})\right.\nonumber\\
&&\hspace{4.8cm}\left.+I_4(m^2_{\tilde{e}_i},m^2_{\tilde{e}_j},m^2_{u_k},m^2_{u_m})+I_4(m^2_{e_i},m^2_{e_j},m^2_{\tilde{u}^k_L},m^2_{\tilde{u}^m_L})\right],\\
\widetilde{C}^{\lambda''}_1&=&\frac{1}{32\pi^2}\sum_{i,j}\lambda''_{i21}\lambda''^*_{i31}\lambda''_{j21}\lambda''^*_{j31}
\left[I_4(m^2_{d},m^2_{d},m^2_{\tilde{u}^i_R},m^2_{\tilde{u}^j_R})+I_4(m^2_{\tilde{d}_R},m^2_{\tilde{d}_R},m^2_{u_i},m^2_{u_j})\right],\\
C^{TL}_4&=&\sum_i\frac{\lambda'_{i32}\lambda'^*_{i23}}{m^2_{\tilde{\nu}_i}},\\
C^{\lambda'W}_4&=&\frac{G_F}{4\sqrt{2}\pi^2}\lambda'_{ik2}\lambda'^*_{ip3}V^*_{u_ps}V_{u_kb}F(m^2_{u_k}/m^2_{\tilde{l}_i}),\label{C4lpW}\\
C^{\lambda''W}_4&=&-\frac{\alpha}{4\pi\mbox{sin}^2\theta_W}\sum_{i,j}\lambda''_{i21}\lambda''^*_{j31}V^*_{u_is}V_{u_jb}m_{u_i}m_{u_j}
J_4(m^2_{\tilde{d}_R},M^2_W,m^2_{u_i},m^2_{u_j}),\label{C4lppW}\\
C^{\lambda'}_5&=&-\frac{1}{32\pi^2}\sum_{i,j,k,m}\lambda'^*_{i2k}\lambda'_{j3k}\lambda'_{im2}\lambda'^*_{jm3}
\left[I_4(m^2_{\tilde{\nu}_i},m^2_{\tilde{\nu}_j},m^2_{d_k},m^2_{d_m})+I_4(m^2_{\nu_i},m^2_{\nu_j},m^2_{\tilde{d}^k_R},m^2_{\tilde{d}^m_R})\right],\\
C^{\lambda''W}_5&=&\frac{\alpha}{4\pi\mbox{sin}^2\theta_W}\sum_{i,j}\lambda''_{i21}\lambda''^*_{j31}V^*_{u_is}V_{u_jb}m_{u_i}m_{u_j}
J_4(m^2_{\tilde{d}_R},M^2_W,m^2_{u_i},m^2_{u_j}), \label{Crpv}
\end{eqnarray}}
where the functions $I_4(m^2_1,m^2_2,m^2_3,m^2_4)$ and
$J_4(m^2_1,m^2_2,m^2_3,m^2_4)$ are defined in Ref.
\cite{deCarlos:1996yh}, $F(x)$ is $I(x)$ for $p=k$ as well as
 $L(x)$ for $p\neq k$, and the definitions  of $I(x)$ and $L(x)$ can
be found in Ref. \cite{Saha:2003tq}.

In terms of the effective Hamiltonian given in Eq. (\ref{Eq.Heff}),
$M^s_{12}$ reads
\begin{eqnarray}
M_{12}^s=\frac{\langle B^0_s|\mathcal{H}_{eff}(\Delta B=\Delta
S=2)|\bar{B}^0_s\rangle}{2m_{B_s}}.
\end{eqnarray}

In the SM, the off-diagonal element of the decay
  matrix $\Gamma^{s,SM}_{12}$  may be written as \cite{Benekembs}
\begin{eqnarray}
\Gamma^{s,SM}_{12}=-\frac{G^2_Fm^2_b}{8\pi
M_{B_s}}(V_{cs}V^*_{cb})^2\left[G(x_c)\langle
B^0_s|Q_1|\bar{B}^0_s\rangle+G_2(x_c)\langle
B^0_s|Q_2|\bar{B}^0_s\rangle+\sqrt{1-4x_c}\hat{\delta}_{1/m}\right],
\end{eqnarray}
where $x_c=m_c^2/m_b^2$, $G(x_c)=0.030$ and $G_2(x_c)=-0.937$ at the
$m_b$ scale \cite{Benekembs}, and the $1/m_b$ corrections
$\hat{\delta}_{1/m}$ are given in Ref. \cite{1/mbcorrections}. It's
important to note that NP can significantly affect $M^s_{12}$, but
not $\Gamma^s_{12}$, which is dominated by the CKM favored $b\to
c\bar{c}s$ tree-level decays. Hence
$\Gamma^s_{12}=\Gamma^{s,SM}_{12}$ holds as a good approximation
\cite{Gamma12,Lenz:2006hd}.

In this work, besides the CP violating phase $\phi_s^{J/\psi\phi}$,
the experimental bounds of the following quantities will be
considered.
\begin{itemize}

\item  The $B_s$ mass difference:
$ \Delta M_s= 2\left|M_{12}^{s}\right|.$

\item The $B_s$ width difference \cite{Grossman}:
$\Delta\Gamma_s=\frac{4|Re(M^s_{12}\Gamma^{s*}_{12})|}{\Delta M_s}
\approx2|\Gamma^s_{12}|\mbox{cos}\phi_s.$

\item
%0605182,0612167
The semileptonic CP asymmetry in $B_s$ decays \cite{ASL1,ASL2}:
$A^s_{SL}=\mbox{Im}\left(\frac{\Gamma^s_{12}}{M^s_{12}}\right)
%=\frac{|\Gamma^s_{12}|}{|M^s_{12}|}~\mbox{sin}\phi_s
=\frac{\Delta
\Gamma_s}{\Delta M_s}~\mbox{tan}\phi_s.$

\end{itemize}

% 0604112
% The CP asymmetry in $B_s$ decays into final CP eigenstates such as $\psi\phi$:
% \begin{eqnarray}
% S_{\psi\phi}=\mbox{sin}\phi_s.
% \end{eqnarray}

\section{Numerical results and discussions}
In this section, we summarize our numerical results and analysis of
$\spur{L}$ as well as  $\spur{B}$ coupling  effects in
$B_s^0-\bar{B}_s^0$ mixing. The theoretical input parameters used in
our work are collected in Table \ref{INPUT}.
\begin{table}[t]
\caption{ Values of the theoretical input parameters. To be
conservative, we use all theoretical input parameters at 95\% C.L.
in our numerical results.
%hep-lat/0110117
The $B^s_i$ parameters are taken from Ref. \cite{Bparameter} in the
RI/MOM  scheme, where the first error is the statistical one and the
second error is the systematic one. }
\begin{center}
\begin{tabular}{lc}\hline\hline
 $m_W=80.398\pm
0.025~GeV,~~~m_{B_s}=5.3663\pm 0.0006~GeV,$&\\
$\overline{m}_b(\overline{m}_b)=4.20^{+0.17}_{-0.07}~GeV,~~~\overline{m}_s(2GeV)=0.105^{+0.025}_{-0.035}~GeV,$&\\
$m_t=171.3^{+2.1}_{-1.6}~GeV,~~~m_b=4.85\pm0.15~GeV$.&\cite{PDG}\\\hline
$A=0.810\pm0.013,~\lambda=0.2259\pm0.0016,~\bar{\rho}=0.177\pm0.044,~\bar{\eta}=0.360\pm0.031$.&\cite{UTfit}\\
\hline $\eta_{2B}=0.55\pm0.01.$&\cite{eta2B}\\\hline
$f_{B_s}=0.230\pm0.030~GeV.$&\cite{decayconstant}\\\hline
$B_{1}^{(s)}(m_b)=0.86(2)\left(^{+5}_{-4}\right),~~B_{2}^{(s)}(m_b)=0.83(2)(4),~~B_{3}^{(s)}(m_b)=1.03(4)(9)$,&\\
$B_{4}^{(s)}(m_b)=1.17(2)\left(^{+5}_{-7}\right),~~B_{5}^{(s)}(m_b)=1.94(3)\left(^{+23}_{-7}\right)$.&\cite{Bparameter}\\\hline
\end{tabular}
\end{center}\label{INPUT}
\end{table}
In our numerical results, we use the  theoretical input parameters
and the experimental constraints at 95\% C.L.. For the general case,
we consider the $\spur{R}_p$ coupling constants are complex, the
phases of the $\spur{R}_p$ coupling products are varied from $-\pi$
to $\pi$, while the moduli of the coupling products are assumed to
be only positive. When we study the $\spur{R}_p$ effects, we
consider only one kind of the $\spur{R}_p$ coupling contributions at
one time, neglecting the interferences between different kinds of
the $\spur{R}_p$ coupling products, but keeping their interferences
with the SM contributions. In addition, we assume the masses of
sfermions are 500 GeV.

% For other values of the sfermion masses, the bounds on the couplings
% in this paper can be easily obtained by scaling them by factor
% $\tilde{f}^2\equiv(\frac{m_{\tilde{f}}}{500~\rm{GeV}})^2$.

The following experimental data at 95\% C.L. are used to constrain
relevant $\spur{R}_p$ couplings
\cite{exphis,Aaltonen:2007he,:2008fj,Abazov:2010hv}
%new data for AsSL from arxiv:0904.3907
\begin{eqnarray}
\phi_s^{J/\psi\phi} &\in& [0.20, 2.84]~ (\mbox{at}~ 95\%~\mbox{C.L.}),\label{exp.phis}\\
\Delta M_s&=&17.77\pm0.12,\label{exp.Dms}\\
 \Delta\Gamma_s&=&0.19\pm0.07,\label{exp.Dgamma}\\
A^s_{SL}&=&(1.46\pm0.75)\times10^{-2}.\label{exp.As}
\end{eqnarray}

\subsection{ $\spur{L}$  couplings}

Now we discuss the constrains on the  $\spur{L}$ coupling spaces. By
using the experimental data and the theoretical input parameters at
95\% C.L., which are given in Eqs. (\ref{exp.phis}-\ref{exp.As}) and
Table \ref{INPUT}, respectively, we  constrain the allowed ranges of
relevant $\spur{L}$ couplings.

\begin{figure}[t]
\begin{center}
\includegraphics[scale=0.8]{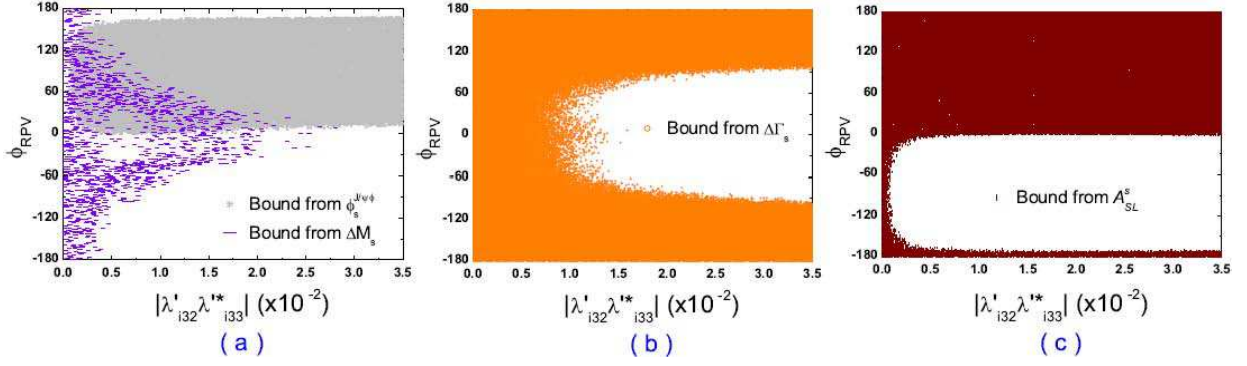}
\end{center}
\vspace{-0.9cm} \caption{ Allowed parameter space for
$\lambda'_{i32}\lambda'^*_{i33}$ constrained
 by $\phi_s^{J/\psi\phi}$ (light gray), $\Delta M_s$ (violet), $\Delta\Gamma_s$ (orange) and
 $A^s_{SL}$ (wine), respectively.  The $\spur{R}_p$ weak phases $\phi_{RPV}$ are given in degree. }
 \label{fig.sep}
\end{figure}

First of all, we take $\lambda'_{i32}\lambda'^*_{i33}$ couplings,
which arise from the $\lambda'-W$ box diagram such as Fig.
\ref{RPVdiagram}(e), as an example to illuminate the bounds from
different observables in Eqs. (\ref{exp.phis}-\ref{exp.As}). Fig.
\ref{fig.sep} displays the allowed spaces for
$\lambda'_{i32}\lambda'^*_{i33}$ constrained
 by the experimental data of $\phi_s^{J/\psi\phi}$, $\Delta M_s$, $\Delta\Gamma_s$ and
 $A^s_{SL}$.
In Fig. \ref{fig.sep}, we only show the
 regions of $|\lambda'_{i32}\lambda'^*_{i33}|$ belongs to
 $[0,3.5\times10^{-2}]$,
and  the regions of
 $|\lambda'_{i32}\lambda'^*_{i33}|>3.5\times10^{-2}$  is still
 allowed
 for the bounds from $\phi_s^{J/\psi\phi}$, $\Delta\Gamma_s$ and
 $A^s_{SL}$.
 The light gray region in Fig. \ref{fig.sep} (a) shows the
 constrained space from $\phi_s^{J/\psi\phi}$, and we can see that current data
 of $\phi_s^{J/\psi\phi}$ at 95\% C.L. give quite strong constraint on  the $\spur{R}_p$ weak phases of
 $\lambda'_{i32}\lambda'^*_{i33}$. Moreover, the lower limits of
 $|\lambda'_{i32}\lambda'^*_{i33}|$ are also constrained by
 $\phi_s^{J/\psi\phi}$
  since its data  at 95\% C.L. are not consistent with its SM value.
The violet region in Fig. \ref{fig.sep} (a) displays the constrained
space from $\Delta M_s$, and we see that current data of $\Delta
M_s$ at 95\% C.L. could obviously constrain the $\spur{R}_p$ weak
phases as well as the upper limits of
$|\lambda'_{i32}\lambda'^*_{i33}|$.
The orange region in Fig. \ref{fig.sep} (b) is  the constrained
space from $\Delta\Gamma_s$, and we see that  $\Delta\Gamma_s$ gives
the bound on the $\spur{R}_p$ weak phases when
$|\lambda'_{i32}\lambda'^*_{i33}|>1.4\times10^{-2}$.
The wine region in Fig. \ref{fig.sep} (c) shows the constrained
space from  $A^s_{SL}$, and we see that  whole region for
$\phi_{RPV}>0$ and some region for $\phi_{RPV}<0$ are allowed. We
find that the bound from $A^s_{SL}$ displayed in Fig. \ref{fig.sep}
(c) is weaker than one from $\phi_s^{J/\psi\phi}$ displayed in Fig.
\ref{fig.sep} (a), therefore $A^s_{SL}$ does not give any useful
constraint if we consider all experimental data given in Eq.
(\ref{exp.As}) to constrain $\lambda'_{i32}\lambda'^*_{i33}$
couplings.
From Fig. \ref{fig.sep}, we know that, if we consider the
experimental bounds of $\phi_s^{J/\psi\phi}$, $\Delta M_s$ and
$\Delta\Gamma_s$ at the same time, the upper limits of
$|\lambda'_{i32}\lambda'^*_{i33}|$ are constrained by $\Delta M_s$
and $\Delta\Gamma_s$, the lower limits of
$|\lambda'_{i32}\lambda'^*_{i33}|$ are  constrained by
$\phi_s^{J/\psi\phi}$, both upper and lower limits of the
$\spur{R}_p$ weak phase are constrained by $\Delta M_s$ and
$\phi_s^{J/\psi\phi}$.

Next, using the experimental data in
Eqs.(\ref{exp.phis}-\ref{exp.As}), we give the constrained spaces of
relevant $\spur{L}$ couplings.
\begin{figure}[ht]
\begin{center}
\includegraphics[scale=0.9]{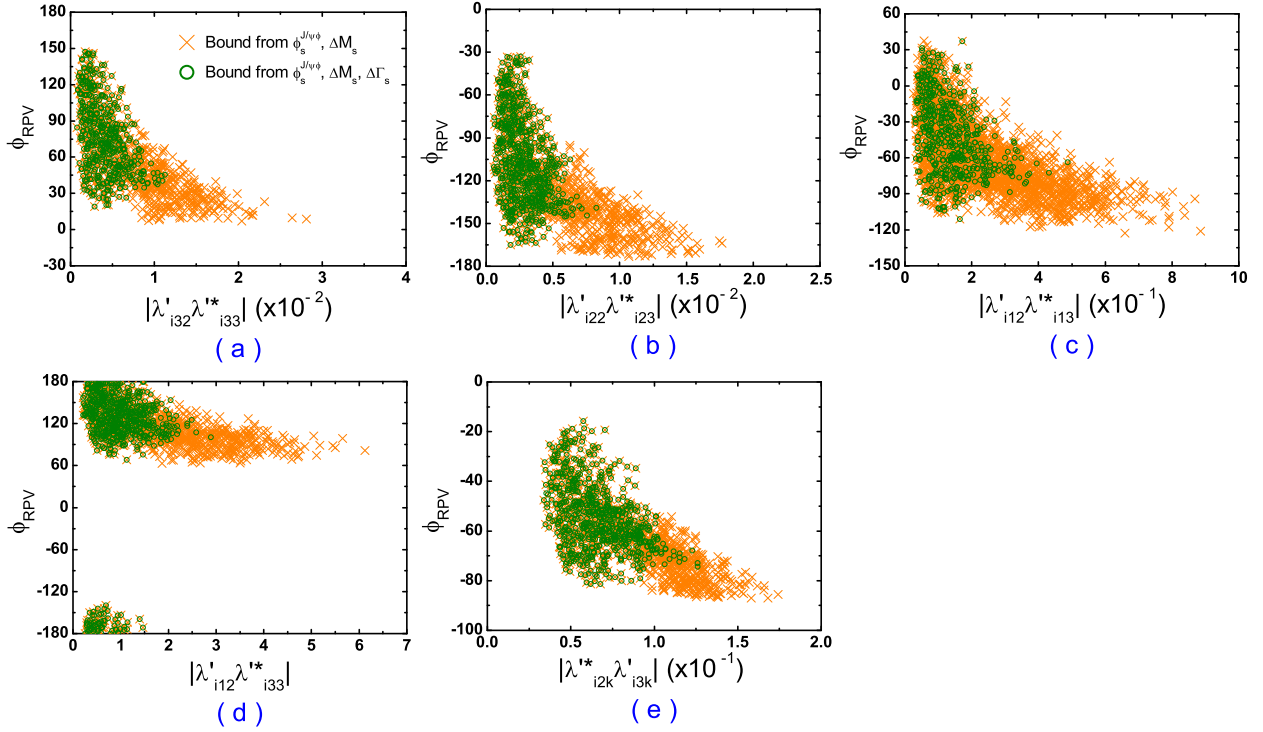}
\end{center}
\vspace{-0.9cm}
 \caption{ The constrained parameter spaces for some relevant $\spur{L}$ couplings.}
 \label{fig.LP}
\end{figure}
Fig. \ref{fig.LP} shows the allowed spaces which arise from the
$\lambda'-W$ box diagram as displayed in Fig. \ref{RPVdiagram}(e).
Other constrained parameter spaces of $\spur{L}$ couplings have not
been on show, since their allowed $\spur{R}_p$ weak phases have
similar allowed regions to one of the plots in Fig. \ref{fig.LP}.
From Fig. \ref{fig.LP}, we see that current experimental data of
$\phi_s^{J/\psi\phi}$ and $\Delta M_s$ give very strong bounds on
both moduli and phases of all relevant $\spur{L}$ coupling products,
and  the upper limits of moduli  of all $\spur{L}$ coupling products
 are further restricted by
$\Delta\Gamma_s$.

Now we describe the correlation of the $\spur{L}$ coupling phases as
follows.
The $\spur{L}$ couplings $\lambda'_{i22}\lambda'^*_{i33}$, which
arise from the $\lambda'-W$ box diagram,  have similar allowed
phases as shown in Fig. \ref{fig.LP}(a).
The couplings
$\lambda'^*_{i2k}\lambda'_{i3k}\lambda'_{ik2}\lambda'^*_{ik3}$,
which are from the $\lambda'$ box diagrams such as Fig.
\ref{RPVdiagram}(c), also have similar allowed phases as shown in
Fig. \ref{fig.LP}(a).
$\lambda'_{i32}\lambda'^*_{i23}$ from tree-level diagram shown in
Fig.
 \ref{RPVdiagram}(b) have similar
allowed regions of $\spur{R}_p$ phases to the region in Fig.
\ref{fig.LP}(b).
$\lambda'_{i22}\lambda'^*_{i13}$, $\lambda'_{i22}\lambda'^*_{i13}$,
$\lambda'_{i32}\lambda'^*_{i13}$ and
$\lambda'_{i32}\lambda'^*_{i23}$ from the $\lambda'-W$ box diagram
also have  similar phase regions to one in Fig. \ref{fig.LP}(b).
The phases of $\lambda'_{i12}\lambda'^*_{i23}$ constrained from the
$\lambda'-W$ box diagram have similar region to Fig.
\ref{fig.LP}(c).
$\lambda'_{im2}\lambda'^*_{im3}(m=1,2,3)$ and
$\lambda'^*_{i2k}\lambda'_{i3k}(k=1,2,3)$ are from  the $\lambda'$
box diagrams  such as Fig. \ref{RPVdiagram}(c-d), and the
constrained phases of $\lambda'_{im2}\lambda'^*_{im3}(m=1,2,3)$ have
similar allowed regions to the constrained phases of
$\lambda'^*_{i2k}\lambda'_{i3k}$ displayed in Fig. \ref{fig.LP}(e).

\begin{table}[t]
\caption{ Bounds on moduli of the  relevant $\spur{L}$ coupling
products for 500 GeV sfermions,  and previous bounds are listed for
comparison. The allowed ranges  within the square brackets are
obtained from the experimental constraints given in Eqs.
(\ref{exp.phis}-\ref{exp.As}) and the theoretical input parameters
listed in Table \ref{INPUT} at 95\% C.L..
 ``$b$" denotes that the
couplings are constrained from the $\lambda'$ box diagrams such as
Fig. \ref{RPVdiagram}(c-d), ``$b'$" denotes that the couplings are
constrained from the $\lambda'-W$ box diagram as Fig.
\ref{RPVdiagram}(e), and ``$t$" denotes that the couplings are
bounded from the tree level diagram as Fig. \ref{RPVdiagram}(b).
(The similar signs are used in Table \ref{Tab.RPVBpp}).  } {\small
\begin{center}
\begin{tabular}{clccl}\hline\hline Line No.&
Couplings & From $\phi_s^{J/\psi\phi}$,$\Delta M_s$ &From $\phi_s^{J/\psi\phi}$,$\Delta M_s$,$\Delta\Gamma_s$ & Previous Bounds\\
1&$|\lambda'_{i32}\lambda'^*_{i33}|(\times10^{-1})^b$&$[0.30,1.91]$&$[0.30,1.39]$&$<8.75$~\cite{deCarlos:1996yh}\\
2&$|\lambda'_{i32}\lambda'^*_{i33}|(\times10^{-2})^{b'}$&$[0.08,2.81]$&$[0.08,1.13]$&$<3.9$~\cite{Nandi:2006qe}\\
3&$|\lambda'_{i22}\lambda'^*_{i23}|(\times10^{-1})^b$&$[0.29,1.74]$&$[0.29,1.26]$&$<8.75$~\cite{deCarlos:1996yh}\\
4&$|\lambda'_{i22}\lambda'^*_{i23}|(\times10^{-2})^{b'}$&$[0.06,1.87]$&$[0.06,1.06]$&$<3.1$ \cite{Nandi:2006qe}\\
5&$|\lambda'_{i12}\lambda'^*_{i13}|(\times10^{-1})^b$&$[0.29,1.74]$&$[0.29,1.26]$&$<8.75$~\cite{deCarlos:1996yh}\\
6&$|\lambda'_{i12}\lambda'^*_{i13}|(\times10^{-1})^{b'}$&$[0.26,8.85]$&$[0.26,4.88]$&$<1.45$
\cite{Nandi:2006qe}\\
7&$|\lambda'_{i32}\lambda'^*_{i23}|(\times10^{-5})^{t}$&$[0.13,4.07]$&$[0.13,1.84]$&$<3.5$ \cite{Wang:2006xha}\\
8&$|\lambda'_{i32}\lambda'^*_{i23}|(\times10^{-4})^{b'}$&$[0.28,9.13]$&$[0.28,4.57]$&$<10.5$ \cite{Nandi:2006qe}\\
9&$|\lambda'_{i22}\lambda'^*_{i33}|(\times10^{-1})^{b'}$&$[0.15,5.22]$&$[0.15,2.71]$&$<5.9$ \cite{Nandi:2006qe}\\
10&$|\lambda'_{i22}\lambda'^*_{i13}|(\times10^{-2})^{b'}$&$[0.24,7.87]$&$[0.24,3.65]$&$<4.9$ \cite{Nandi:2006qe}\\
11&$|\lambda'_{i12}\lambda'^*_{i23}|(\times10^{-1})^{b'}$&$[0.07,2.29]$&$[0.07,1.38]$&$<1.0$ \cite{Nandi:2006qe}\\
12&$|\lambda'_{i32}\lambda'^*_{i13}|(\times10^{-3})^{b'}$&$[0.13,4.10]$&$[0.13,1.80]$&$<4.6$ \cite{Nandi:2006qe}\\
13&$|\lambda'_{i12}\lambda'^*_{i33}|^{b'}$&$[0.21,6.13]$&$[0.21,2.89]$&$<3.1$ \cite{Nandi:2006qe}\\
14&$|\lambda'_{i23}\lambda'^*_{i33}|(\times10^{-1})^b$&$[0.29,1.74]$&$[0.29,1.26]$&$<2.0$ \cite{Nandi:2006qe,deCarlos:1996yh}\\
15&$|\lambda'_{i22}\lambda'^*_{i32}|(\times10^{-1})^b$&$[0.29,1.74]$&$[0.29,1.26]$&$<2.0$ \cite{Nandi:2006qe,deCarlos:1996yh}\\
16&$|\lambda'_{i21}\lambda'^*_{i31}|(\times10^{-1})^b$&$[0.29,1.74]$&$[0.29,1.26]$&$<2.0$
\cite{Nandi:2006qe,deCarlos:1996yh} \\
17&$|\lambda'^*_{i2k}\lambda'_{i3k}\lambda'_{ik2}\lambda'^*_{ik3}|(\times10^{-2})^{b}$&$[0.06,2.43]$&$[0.06,1.23]$&\\
\hline
\end{tabular}
\end{center}\label{Tab.RPVLp}}
\end{table}

The relevant numerical bounds on moduli of $\spur{L}$  coupling
products  are  summarized in Table \ref{Tab.RPVLp}. Previous bounds
are also listed for comparison. We present some remarks on the
moduli of all relevant $\spur{L}$  coupling products:
\begin{itemize}

\item Almost all bounds on the moduli of $\spur{L}$ coupling
products from current 95\% C.L. data of $\Delta M_s$,
$\phi_s^{J/\psi\phi}$ and $\Delta\Gamma_s$ are stronger than
previous ones in Refs. \cite{Nandi:2006qe,Wang:2006xha}, which are
obtained only from the  68\% C.L.  data of $\Delta M_s$.  $\spur{L}$
couplings that may contribute to $B_s^0-\bar{B}_s^0$ mixing also
affect various decays, relevant $\spur{L}$ coupling effects in the
decays have already been studied in Refs.
\cite{Nandi:2006qe,deCarlos:1996yh,Yang:2005es,Xu:2006vk}, and some
moduli of them may have better bounds from the decays.

% However, some direct product bounds, which occur from neutrino
% constraints with no mixing scenario, are indeed tight
% \cite{Allanach:1999ic}.

\item $\lambda'_{ik2}\lambda'^*_{ik3}(k=1,2,3)$ couplings arise in both the $\lambda'$ box
diagrams and the $\lambda'-W$ box diagram. In this work, we only
consider one kind of diagrams at once.  For
$|\lambda'_{ik2}\lambda'^*_{ik3}|(k=2,3)$, as listed in the first
four lines of Table \ref{Tab.RPVLp}, the contributions of these
couplings to the $\lambda'$  box diagrams are smaller than ones to
the $\lambda'-W$ box diagram, and there are more than
 one order difference between two kinds of diagrams. Therefore, the stronger bounds on
$|\lambda'_{ik2}\lambda'^*_{ik3}|(k=2,3)$ are constrained from the
$\lambda'-W$ box diagram. If we consider these two kinds of diagrams
at the same time, the bounds on
$|\lambda'_{ik2}\lambda'^*_{ik3}|(k=2,3)$ are closed  to ones only
from the $\lambda'-W$  diagram.
For $|\lambda'_{i12}\lambda'^*_{i13}|$,  the contributions  from the
$\lambda'$ box diagrams are sensitive to sfermion masses, but  the
contributions  from the $\lambda'-W$ box diagram are not sensitive
to sfermion masses. For 500 GeV sfermion masses, the couplings
raised in both the $\lambda'$ box diagrams and the $\lambda'-W$ box
diagram give  similar contributions.

\item The contributions of  $\lambda'_{ik2}\lambda'^*_{ik3}$ from the
$\lambda'-W$ box diagram are proportional to the CKM matrix elements
$V^*_{u_ks}V_{u_kb}$ and the function
$I(m^2_{u_k}/m^2_{\tilde{l}_i})$.  The bound differences between
$|\lambda'_{i32}\lambda'^*_{i33}|$ and
$|\lambda'_{ik2}\lambda'^*_{ik3}|(k=1,2)$ come from  the CKM matrix
elements and internal quark mass $m_{u_k}$.
The bound differences between $|\lambda'_{i12}\lambda'^*_{i13}|$ and
$|\lambda'_{i22}\lambda'^*_{i23}|$ mainly come from  the CKM matrix
elements  since $I(m^2_{c}/m^2_{\tilde{l}_i})\approx
I(m^2_{u}/m^2_{\tilde{l}_i})$. For this reason, as listed in  Lines
4 and 6 of Table \ref{Tab.RPVLp},
$|\lambda'_{i12}\lambda'^*_{i13}|/|\lambda'_{i22}\lambda'^*_{i23}|\approx
|V^*_{cs}V_{cb}|/|V^*_{us}V_{ub}|\approx46$.

\item  $\lambda'_{i32}\lambda'^*_{i23}$ couplings arise in
both the tree level diagram  and the $\lambda'-W$ box diagram. Compared bounds listed in Lines 7-8 of Table
\ref{Tab.RPVLp}, we can see that  the stronger bounds on
$|\lambda'_{i32}\lambda'^*_{i23}|$
 come from the tree level contributions.

\item $\lambda'_{ik2}\lambda'^*_{ip3}(k\neq p)$ couplings arise in the $\lambda'-W$ box diagram.
 As given in Eq.
(\ref{C4lpW}), their contributions are proportional to the
corresponding CKM matrix elements $V^*_{u_ps}V_{u_kb}$ and the
function $F(m^2_{u_k}/m^2_{\tilde{l}_i})$. Their bounds are listed
in
 Lines 8-13 of Table \ref{Tab.RPVLp}, and
the bound differences  between $|\lambda'_{ik2}\lambda'^*_{ip3}|$
and $|\lambda'_{ip2}\lambda'^*_{ik3}|$  are due to the different CKM
matrix elements.

\end{itemize}

%\clearpage

\subsection{ $\spur{B}$  couplings}
\label{Bsubsection}

Now we turn to discuss $\spur{B}$ couplings. As given in Eqs.
(\ref{C4lppW}-\ref{Crpv}), $\spur{B}$ couplings
$\lambda''_{i21}\lambda''^*_{j31}(i,j=1,2,3)$ from the $\lambda''-W$
box diagram such as Fig. \ref{RPVdiagram}(g) are suppressed by the
CKM matrix elements $V^*_{u_is}V_{u_jb}$ and the quark masses
$m_{u_i}m_{u_j}$. Therefore, we don't consider the $\spur{B}$
couplings of $\lambda''_{i21}\lambda''^*_{j31}(i~ \mbox{or}~ j=1)$
in this work since they  are significantly suppressed by $m_u$,
$V_{us}$ and $V_{ub}$. We can see later the bounds on moduli of
$\lambda''_{i21}\lambda''^*_{j31}(i,j\neq1)$ couplings from the
$\lambda''-W$ box diagram of Fig. \ref{RPVdiagram}(g) are very weak
for this reason.

\begin{figure}[t]
\begin{center}
\includegraphics[scale=1.5]{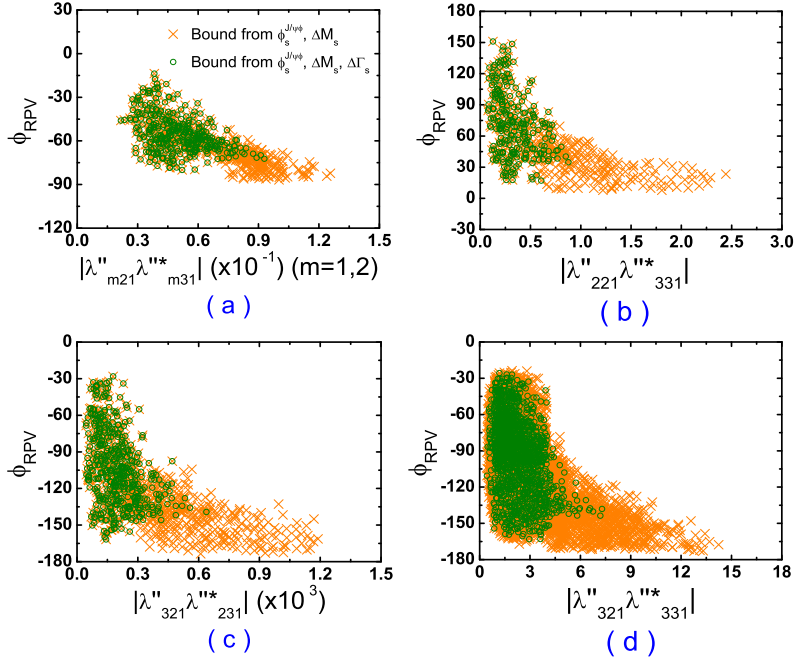}
\end{center}
 \caption{ The constrained parameter spaces for some relevant $\spur{B}$ couplings.}
 \label{fig.LPP}
\end{figure}
In Fig. \ref{fig.LPP}, we show some constrained parameter spaces of
$\spur{B}$ couplings from the experimental data given in Eqs.
(\ref{exp.phis}-\ref{exp.As}).
Fig. \ref{fig.LPP}(a) displays the constrained spaces of
$\lambda''_{m21}\lambda''^*_{m31}(m=1,2)$ couplings, which are from
the $\lambda''$ box diagrams such as  Fig. \ref{RPVdiagram}(f).
The constrained spaces of $\lambda''_{221}\lambda''^*_{331}$,
$\lambda''_{321}\lambda''^*_{231}$ and
$\lambda''_{321}\lambda''^*_{331}$, which present in the
$\lambda''-W$ box diagram as Fig. \ref{RPVdiagram}(g), are displayed
in Fig. \ref{fig.LPP}(b-d).
Other two relevant couplings $\lambda''_{321}\lambda''^*_{331}$ and
$\lambda''_{221}\lambda''^*_{231}$, which arise in the $\lambda''$
box diagrams,  have not been shown in Fig. \ref{fig.LPP}. The
allowed phase of $\lambda''_{321}\lambda''^*_{331}$  is similar to
one in Fig. \ref{fig.LPP}(a), and the allowed phase of
 $\lambda''_{221}\lambda''^*_{231}$ is
similar to one in Fig. \ref{fig.LPP}(b).
We can see that current 95\% C.L. experimental data of
$B_s^0-\bar{B}_s^0$ mixing give very strong bounds on relevant
$\spur{B}$ phases. However, they do not constrain some moduli a lot.

\begin{table}[tb]
\caption{ Bounds on moduli of the  relevant $\spur{B}$ coupling
products for 500 GeV sfermions. The allowed ranges  within the
square brackets are obtained from the experimental data given in
Eqs. (\ref{exp.phis}-\ref{exp.As}) and the theoretical input
parameters listed in Table \ref{INPUT} at 95\% C.L.. }
\vspace{0.3cm}
\begin{center}
\begin{tabular}{clcc}\hline\hline Line No.&
Couplings & From $\phi_s^{J/\psi\phi}$,$\Delta M_s$ &From $\phi_s^{J/\psi\phi}$,$\Delta M_s$,$\Delta\Gamma_s$ \\
1&$|\lambda''_{321}\lambda''^*_{331}|(\times10^{-1})^b$&$[0.22,1.41]$&$[0.22,0.96]$ \\
2&$|\lambda''_{321}\lambda''^*_{331}|(\times10^{-1})^{b'}$&$[0.43,14.23]$&$[0.43,7.29]$\\
3&$|\lambda''_{221}\lambda''^*_{231}|(\times10^{-1})^b$&$[0.19,1.26]$&$[0.19,0.93]$\\
4&$|\lambda''_{221}\lambda''^*_{231}|(\times10^{2})^{b'}$&$[0.10,3.47]$&$[0.10,1.66]$\\
5&$|\lambda''_{121}\lambda''^*_{131}|(\times10^{-1})^b$&$[0.19,1.26]$&$[0.19,0.93]$\\
6&$|\lambda''_{321}\lambda''^*_{231}|(\times10^{3})^{b'}$&$[0.04,1.19]$&$[0.04,0.64]$\\
7&$|\lambda''_{221}\lambda''^*_{331}|^{b'}$&$[0.07,2.44]$&$[0.07,0.92]$\\
\hline
\end{tabular}
\end{center}\label{Tab.RPVBpp}
\end{table}

Our bounds on the moduli of relevant $\spur{B}$ coupling products
are listed in Table \ref{Tab.RPVBpp}. Table \ref{Tab.RPVBpp} shows
us that the upper limits of all relevant moduli  of $\spur{B}$
coupling products are further restricted by $\Delta\Gamma_s$. From
the first six lines of this table, we easily see that the
$\lambda''$ box diagrams give the dominant contributions to
$B_s^0-\bar{B}_s^0$ mixing. The bounds on
$|\lambda''_{i21}\lambda''^*_{i31}|$ from the $\lambda''$ box
diagrams are much stronger than ones from the $\lambda''-W$ box
diagram, since the contributions of the $\lambda''-W$ box diagram
are suppressed by internal up-type quark masses and relevant CKM
matrix elements. Therefore, if we consider the contributions from
the $\lambda''$ box diagrams and the $\lambda''-W$ box diagram at
the same time, the bound on $|\lambda''_{i21}\lambda''^*_{i31}|$ is
close to the constraint only from the $\lambda''$ box diagrams.
As listed  in the last line of Table \ref{Tab.RPVBpp},
$|\lambda''_{321}\lambda''^*_{231}|$ has  very weak constraints from
current data, since the $\lambda''_{321}\lambda''^*_{231}$
contributions are suppressed by $m_c$ and $V_{ts}V_{cb}$.

The relevant bounds of the $\spur{B}$ coupling products have already
been obtained in Refs.
\cite{deCarlos:1996yh,Chakraverty:2000rm,Yang:2005es,Chemtob:2004xr},
and we summarize them with 500 GeV sfermion mass as follows.
\begin{itemize}

\item $|\lambda''_{i21}\lambda''^*_{i31}|<1.0\times10^{-1}$ from
$B^+\to \bar{K}^0\pi^+$ decay \cite{Chemtob:2004xr}.

\item $|\lambda''_{i21}\lambda''^*_{i31}|<1.54\times10^{-1}$ from
$\frac{\Gamma(B^+\to \bar{K}^0\pi^+)}{\Gamma(B^+\to J/\psi K^+)}$
\cite{deCarlos:1996yh}.

\item $|\lambda''_{i21}\lambda''^*_{i31}|\in [9.3\times
10^{-3},1.2\times 10^{-1}]\cup[3.4\times 10^{-1},4.0\times 10^{-1}]$
from $B^+\to \bar{K}^0\pi^+,K^+\pi^0$ \cite{Yang:2005es}.

\item $|\lambda''_{i2k}\lambda''^*_{i3k}|<4$ from $B\to K^* \gamma$
decays \cite{Chemtob:2004xr}.

\item $|\lambda''_{32k}\lambda''^*_{33k}|<8.75$ from $\mathcal{B}(B\to
X_s\gamma)$ \cite{Chakraverty:2000rm}.

\end{itemize}
Comparing our results with the existing ones listed above, our upper
limits of $|\lambda''_{i21}\lambda''^*_{i31}|(i=1,2,3)$ from the
$\lambda''$ box diagrams listed in Lines 1,3,5 of Table
\ref{Tab.RPVBpp}  are a little stronger than the existing ones. In
addition, the lower limits are also obtained from
$B_s^0-\bar{B}_s^0$ mixing since some relevant data are not
consistent with the SM predictions at 95\% C.L. As for
$\lambda''_{321}\lambda''^*_{231}$ and
$\lambda''_{221}\lambda''^*_{331}$ coupling products, their bounds
are derived for the first time in this work.

\section{Summary}
The flavor changing processes in the $b-s$ sector are sensitive to
probing  of NP beyond the SM because they have the least constraint
in current experiment aspect.  Recent measurements of the CP
violating phase by the D{\O} and CDF collaborations exclude the SM
predictions at 95\% C.L., and this suggests NP beyond the SM
contributing to $B_s^0-\bar{B}_s^0$ mixing.  Motivated by this, we
have analyzed the constraints imposed on the parameter space of
$\spur{L}$ and $\spur{B}$ contributions to $M^s_{12}$ in
$\spur{R}_p$ SUSY. We have shown that current experimental data in
$B_s^0-\bar{B}_s^0$ mixing can be explained by the $\spur{R}_p$ SUSY
effects. Current data of $\Delta M_s$, $\phi_s^{J/\psi\phi}$ and
$\Delta\Gamma_s$  give quite strong bounds on some moduli and phases
of relevant couplings. And the data of $A^s_{SL}$ doesn't give any
useful constraint  since the bounds from $A^s_{SL}$ are weaker than
ones from $\phi_s^{J/\psi\phi}$.

We first considered  $\spur{L}$ coupling effects in
$B_s^0-\bar{B}_s^0$ mixing. The similar analysis was performed only
from the bound on $\Delta M_s$ in Refs.
\cite{Wang:2006xha,Nandi:2006qe}, in this paper we have used the
current bounds not only on $\Delta M_s$ but also on
$\phi_s^{J/\psi\phi}$ and $\Delta\Gamma_s$  as given in Eqs.
(\ref{exp.phis}-\ref{exp.As}).  We have found that almost all bounds
on the moduli of $\spur{L}$ coupling products from current 95\% C.L.
data of $\Delta M_s$, $\phi_s^{J/\psi\phi}$ and $\Delta\Gamma_s$ are
much stronger than previous ones only from the 68\% C.L. data of
$\Delta M_s$ \cite{Nandi:2006qe,Wang:2006xha}. We also have obtained
quite strong bounds on the $\spur{R}_p$ weak phases of these
coupling products. Noted that some $\spur{L}$ couplings, which may
contribute to $B_s^0-\bar{B}_s^0$ mixing, also affect various
decays, and the moduli of these $\spur{L}$ couplings may still have
better bounds from relevant decays.

For $\spur{B}$ coupling effects in  $B_s^0-\bar{B}_s^0$ mixing, we
studied them for the first time in this work. We have found that our
bounds on the moduli of $\lambda''_{i21}\lambda''^*_{i31}(i=1,2,3)$
from  current data of
 $B_s^0-\bar{B}_s^0$ mixing are stronger than ones from
relevant decays. And we have obtained very strong bounds on the
$\spur{R}_p$ weak phases of
$\lambda''_{i21}\lambda''^*_{i31}(i=1,2,3)$. In addition, the bounds
on the $\spur{B}$ coupling products,
$\lambda''_{321}\lambda''^*_{231}$ and
$\lambda''_{221}\lambda''^*_{331}$,   have been derived for the
first time.

It should be noted that there still are allowed parameters for all
relevant $\spur{L}$ and $\spur{B}$ coupling products if we also add
the bounds of $C_{B_s},~ \phi_{B_s},~\phi^{NP}_s$ and $R_s$ from
Unitarity Triangle analysis in Ref. \cite{Bona:2009tn}. Comparing
with the bounds from $\Delta M_s$, $\phi_s^{J/\psi\phi}$ and
$\Delta\Gamma_s$, we find that, after considering to add the bounds
of $C_{B_s},~ \phi_{B_s},~\phi^{NP}_s$ and $R_s$,  the lower limits
of moduli of relevant $\spur{R}_p$ coupling products will be shrunk
and the ranges of the allowed $\spur{R}_p$ weak phases will be
decreased.
More detailed measurements of relevant observables  at the Tevatron,
the LHC and the B-factories in near future can shrink or reveal the
relevant parameter spaces of relevant $\spur{R}_p$ couplings.

\section*{Acknowledgments}

The authors would like to thank Prof. Alexander Lenz and Dr.
Guennadi Borissov for helpful comments and suggestions on the
manuscript. The work is supported by the National Science Foundation
of P.R. China under contract Nos. 11047145 and 11005088.

\end{document}